\documentclass[aps,prd]{revtex4}
\usepackage[utf8x]{inputenc}
\usepackage{amsmath}
\usepackage{amssymb}
\usepackage{color}
\usepackage{mathrsfs} 

\begin{document}

\title{Lifshitz scalar fields: one-loop renormalization in curved backgrounds}

\author{Diana L. L\'opez Nacir $^{1}$}
\author{Francisco D. Mazzitelli$^{1,2}$}
\author{Leonardo G. Trombetta$^1$}

\affiliation{$^1$ Departamento de F\'\i sica and IFIBA, FCEyN UBA, Facultad de Ciencias Exactas y Naturales,
 Ciudad Universitaria, Pabell\' on I, 1428 Buenos Aires, Argentina}
\affiliation{$^2$ Centro At\'omico Bariloche
Comisi\'on Nacional de Energ\'\i a At\'omica,
R8402AGP Bariloche, Argentina}

\date{\today}

\begin{abstract}
We consider an interacting Lifshitz field with $z=3$ in a curved spacetime.
 We analyze the renormalizability of the theory for interactions of the form $\lambda\phi^n$, with arbitrary even $n$.
 We compute the running of the coupling constants both in the ultraviolet and infrared regimes. We show that the Lorentz-violating terms generate couplings to the spacetime metric that 
are not invariant under general coordinate transformations.  These couplings are not suppressed by the scale of Lorentz violation and therefore survive at low energies. We point out that in these theories, unless the effective mass of the field is many orders of magnitude below the scale of Lorentz violation, the coupling to the four-dimensional Ricci scalar $\xi {}^{(4)}R \phi^2 $ does not receive large quantum corrections $\xi\gg 1$. 
We  argue that quantum corrections involving spatial derivatives of the lapse function (which appear naturally in the so-called healthy extension of the Ho\v{r}ava-Lifshitz theory of gravity) are not generated unless they are already present in the bare Lagrangian.
\end{abstract}


\maketitle

\section{Introduction}

As there is no hint about how the laws of physics are modified at extremely high energies, it is reasonable to explore the possibility of high-energy 
violations of the low-energy symmetries and test how robust the predictions in the infrared are when departures from those symmetries 
in the ultraviolet are considered. In particular, in the last years, there has been a growing interest in field theories where Lorentz 
invariance is explicitly broken at high energies \cite{Liberati}. 

There is an additional motivation to consider these kinds of theories, which comes from gravity. In the theory proposed by Ho\v{r}ava \cite{horava}, 
one introduces a preferred foliation of spacetime and allows for higher spatial derivatives in the Lagrangian. The invariance under general changes
 of coordinates is lost, but the higher spatial derivatives improve the quantum version of the theory.  It becomes a power-counting renormalizable theory
 of gravity if one includes  $2z$ spatial derivatives, with $z\geq 3$.  At low energies, one expects that there will be a region in the parameter space of the theory in which the well-tested predictions of general relativity 
are recovered.

When  matter fields are coupled to Ho\v{r}ava's gravity, in principle, one has two options for  their corresponding bare Lagrangians: 
one could consider that they also have higher spatial derivatives,  breaking explicitly Lorentz invariance in the matter sector, or that
they do not have them. 
The latter  was considered in Ref. \cite{pospelov} with the main goal of making the model phenomenologically viable in the sense that the observational constraints on 
Lorentz-violating effects (which in this case, are induced only by the gravitational couplings) could be naturally satisfied. However, as was shown  in Ref. \cite{DNL-GG-FDM},  
the renormalizability of the theory implies that one must also include higher derivatives in the matter sector. 
Indeed, even for free quantum matter fields in a classical gravitational background, $z\geq 3$ is needed in order to avoid the existence of divergences with more than two
 time-derivatives of the metric \cite{DNL-GG-FDM}. Scalar fields with such higher spatial derivatives, which obviously break Lorentz invariance, are named Lifshitz fields, 
 and are the main subject of the present paper.
 
Various investigations on the properties and applications of Lifshitz fields with $z=3$  can be found in the literature; see  for instance
Refs. \cite{calcagni,mukohyama1,mukohyama2} in the cosmological context, and Refs. \cite{Farakos1,Farakos2} for nonperturbative studies in flat space. There is an important question that deserves to be addressed, which is  whether one can recover the usual Lorentz-invariant field
 theory predictions at low energies or not,  for Lifshitz fields. On this subject, in Refs. \cite{iengo-russo,gomes}, the authors
 considered models with $z=2$ scalar fields (and also models with fermions as a Yukawa like model) in higher spatial dimensions  
with the main goal of analyzing the renormalization group (RG) evolution  of the ``speed-of-light'' couplings $c_i$'s for one and more than one field
 (see also Refs. \cite{iengo-serone,alexandre}). The one-loop renormalized effective potential for a Lifshitz scalar field with $z=2$ at finite
 temperature has been analyzed in Ref. \cite{koreanos}. One motivation for the choice of  $z=2$ and higher dimensions is to obtain nontrivial 
results without having to compute high-loop contributions or diagrams with many legs. The results in Refs. \cite{iengo-russo,gomes} indicate that the recovery
 of the usual Lorentz-invariant theories at low energies is nontrivial, requiring in general a fine-tunning of the parameters in the ultraviolet. 
 The requirement of such fine-tunnings seems to be generic in the sense that even thought the different $c_i$'s can run to the same value in the IR,  in
simple models,  the running is in general too slow to satisfy the phenomenological constraints. However, in Ref. \cite{Donoghue}, the authors have pointed out that a
 faster running can be achieved in some scenarios  containing a large number of hidden fields beyond the standard model ones. Clearly, the studies carried out so far are still not conclusive,
 and  this question should be further addressed.  Although this latter issue goes beyond the scope of the present paper, one of our goals is to move forward 
in this direction by extending nonperturbative RG techniques for $z=3$ Lifshitz fields. 

The analysis of quantum field theory with $z\neq 1$ is therefore of interest for several reasons. On the one hand, if one considers Ho\v{r}ava's gravity as a useful arena to understand different aspects of quantum gravity, it is unavoidable to include these kinds of matter fields. On the other hand, this analysis can be considered as a simplified situation for the study of the renormalizability of Ho\v{r}ava's gravity beyond power-counting analysis. Finally, one could address interesting questions as, for instance, 
 the emergence of the usual Lorentz-invariant theories at low energies. 

With these motivations in mind, in this paper, we  study  the theory of self-interacting quantum Lifshitz fields with $z=3$  in curved backgrounds. We will consider interactions of the form  $\lambda\phi^n$, with arbitrary $n$, that are allowed by power-counting. Looking at the Heisenberg equations for the quantum field operator, we will discuss the one-loop renormalizability of the theory. This turns out to be a rather simple task, since the divergences are those of flat spacetime. It is also possible to compute the running of the coupling constants in the ultraviolet regime. In order to analyze the infrared behavior of the coupling constants, we generalize a method based on the evolution equation for the effective potential \cite{erge,Liao-Polonyi}, to the case of Lifshitz fields.

We consider the quantum fields in a curved background so as to analyze the couplings to the spacetime metric induced by the self-interaction. It is well-known that, in theories with $z=1$,  it is necessary to include couplings  to the curvature of the form  $\xi R\phi^2$ in order to absorb the divergences of the theory. We will see that, although these kinds of terms are not necessary to renormalize for $z=3$, the self-interaction induces finite couplings to both the 3-curvature and the extrinsic curvature. Somewhat surprisingly, these couplings are not suppressed by the scale of Lorentz violation, leaving a footprint of the ultraviolet behavior of the theory in the infrared. We will also analyze whether quantum corrections induce terms containing derivatives of the lapse function or not, as needed in the so-called healthy extension of Ho\v{r}ava's theory \cite{blas}.

\section{Interacting Lifshitz fields in curved backgrounds}

As already mentioned, there are several reasons for considering a $z=3$ scalar field in a curved background, amongst which is to move forward in the understanding 
of quantum effects in the framework of the Ho\v{r}ava-Lifshitz theory.  As was shown in Ref. \cite{anselmi}, a  power-counting analysis  similar to the usual one can be applied for any $z$, provided one
 substitutes the standard scaling dimensions of the operators by their ``weighted scaling dimensions'', i.e. by the dimensions implied by the assignment
$[x]_w = −1$ and $[t]_w = −z$.
A simple power-counting argument  along  the lines of Refs. \cite{visser1,visser2} shows that,  in 
this case, the interacting theory is renormalizable for an interaction of the form $\phi^n$, with $n$ arbitrary.

To review this, let us suppose the action of the scalar field in $D+1$ flat dimensions $S = S_{\textup{free}} + S_{\textup{int}}$, with a free part of the form
\begin{equation}
S_{\textup{free}} = \int \left\{ \dot{\phi}^2 - \phi \left[ m^2 + \Delta + \cdots + \Lambda^{2-2z} \left( \Delta \right)^z \right] \phi \right\} ~ dt ~ d^D x ,
\end{equation}
where $\Lambda$ is a parameter with momentum dimensions that controls the Lorentz-violation scale. For the interaction part of the action, we consider a polynomial in $\phi$ of degree $n_{max}$:
\begin{equation}
S_{\textup{int}} = \int P(\phi) ~ dt ~ d^D x = \int \left\{ \sum_{n=3}^{n_{max}} \frac{\lambda_n}{n!} \phi^n \right\} ~ dt ~ d^D x ,
\end{equation}
where each coupling constant $\lambda_n$ has dimensions
\begin{equation}
[\lambda_n] = M^{D+1-n \frac{(D-1)}{2}} .
\end{equation}

For a generic Feynman diagram with $L$ loops and $I$ internal lines, the superficial degree of divergence \cite{visser1} is
\begin{equation}
\delta = L(D+z)-2Iz = L(D-z)-2z(I-L) .
\label{sup-degree-div}
\end{equation}
As $I \geq L$ in general, for $z=3$, we can expect at most a logarithmic divergence ($\delta = 0$). Setting $z=1$ reduces this expression to the standard result for Lorentz-invariant theories. Considering, in particular, a given diagram with only one type of vertex with $n$ legs, and using Euler's theorem for graphs along with other identities regarding the number of $n$-legged vertices and the number of external lines $E$, the expression \eqref{sup-degree-div} can be rewritten as
\begin{eqnarray}
\delta = \left[ (D+z) \left( 1 - \frac{2}{n} \right) - 2z \right] I + \left( 1 - \frac{E}{n} \right) (D+z) .
\end{eqnarray}
Now, for a given process, the number of external lines $E$ is constant, while the number of loops $L$ and internal lines $I$ increase with every order in perturbation theory. Therefore, $\delta$ will not increase for further corrections if
\begin{equation}
(D+z) \left( 1 - \frac{2}{n} \right) - 2z \leq 0 ,
\end{equation}
which translates to
\begin{eqnarray}
&&n \leq \frac{2(D+z)}{D-z}, ~~~~~~ z<D ,\\
&&n \leq \infty, ~~~~~~~~~~~~~~~\, z\geq D .
\label{n-renorm}
\end{eqnarray}
This shows that for $z\geq D$, the theory is power-counting renormalizable for any power of $\phi$ in the interaction Lagrangian. In the Lorentz-invariant case ($z=1$), the highest power allowed is $n_{max}(z=1) = \frac{2(D+1)}{D-1}$, which gives the standard result $n_{\rm max}=4$ in $3+1$ dimensions.

We will consider a Lorentz noninvariant action in a curved spacetime. Using the Arnowitt-Deser-Misner formalism, the spacetime interval is written as
\begin{equation}
ds^{2}=-N^{2}dt^{2}+g_{ij}(dx^{i}+N^{i}dt)(dx^{j}+N^{j}dt),  \label{ADM}
\end{equation}%
where, as usual, Latin indices refer to the spatial coordinates, $i,j=1,2,3$, $N$ and $N_i$ are the lapse and shift functions, respectively, 
and $g_{ij}$ is the spatial metric. The explicit form of the action is given by
\begin{eqnarray}
S_{\phi} = \int dt d^3 x N \sqrt{g} \Biggl[ \frac{1}{2N^2} \left( \dot{\phi} - N^i \partial_i{\phi} \right)^2 - \frac{1}{2} m^2 \phi^2 - b_1 \partial_i \phi \partial^i \phi - \frac{b_2}{\Lambda^2} ( \Delta \phi )^2 + \frac{b_3}{\Lambda^4} \Delta^2 \phi \Delta \phi - \sum_{p=2}^{p_{max}} \frac{\lambda_{2p}}{(2p)!} \phi^{2p} \Biggr] ,
\label{action}
\end{eqnarray}
where the operator $\Delta$ is the spatial Laplacian associated to $g_{ij}$. The sources of Lorentz violation are the coefficients $b_1$, $b_2$ and $b_3$, while, as before,  $\Lambda$ is a parameter with dimensions of mass that sets the scale at which this violation occurs. Furthermore, we included interactions of the type $\phi^{2p}$ up to an arbitrary value of $p_{max} = n_{max}/2$, which are not renormalizable in the Lorentz-invariant case when $n_{max}>4$. Neither the kinetic terms nor the interaction terms considered here are the most general; for instance, as was emphasized in Ref. \cite{blas}, the foliation-preserving-diffeomorphism symmetry of the theory  also allows the inclusion of spatial derivatives of the logarithm of the lapse function $N$. One could also consider interaction terms containing derivatives of $\phi$. We restrict for the moment to this particular choice, and postpone the discussion about more general Lagrangians until Sec. \ref{moregenLag}.

The classical  equation of motion derived from the action \eqref{action} is 
\begin{equation}
N\sqrt{g} \left[ \left( \square_L(x) - m^2 \right) \phi - \sum_{p=2}^{p_{max}} \frac{\lambda_{2p}}{(2p-1)!} \phi^{2p-1} \right] = 0,
\label{EOM}
\end{equation}
where we have defined the operator
\begin{eqnarray}
\square_L(x)_{\dots} &\equiv& \frac{1}{N \sqrt{g}} \Biggl\{-\partial_t \left( \frac{\sqrt{g}}{N} (\partial_t - N^i \partial_i)_{\dots} \right) + \partial_i \left( \frac{N^i \sqrt{g}}{N} (\partial_t - N^j \partial_j)_{\dots}  \right) + 2 b_1 \partial_i \left( N \sqrt{g} \partial^i_{\dots}  \right)\\ &&- 2 \frac{b_2}{\Lambda^2} \sqrt{g} \Delta \left( N \Delta_{\dots}  \right) + \frac{b_3}{\Lambda^4} \sqrt{g} \Delta \left( N \Delta^2_{\dots}  \right) + \frac{b_3}{\Lambda^4} \sqrt{g} \Delta^2 \left( N \Delta_{\dots}  \right) \Biggr\}.
\label{operator-boxL}
\end{eqnarray}

Proceeding along the same lines as in Ref. \cite{FDM-JPP}, we write the quantum field $\phi$ as its  mean value $\langle \phi \rangle$ plus quantum fluctuations $\hat{\phi}$ around that value, which is
\begin{equation}
\phi = \langle \phi \rangle + \hat{\phi} .
\label{desc-phi}
\end{equation}
For this decomposition to be consistent, the expectation value of the fluctuations must vanish $\langle \hat{\phi} \rangle = 0$. Replacing Eq. \eqref{desc-phi} in the equation of motion \eqref{EOM} and taking its mean value $\langle \dots \rangle$, we are left with the evolution equation for $\phi_0 \equiv \langle \phi \rangle$:
\begin{equation}
\begin{split}
N \sqrt{g} \left[ \left( \square_L(x) - m^2 \right) \phi_0 - \sum_{p=2}^{p_{max}} \sum_{k=0}^{2p-1} \frac{\lambda_{2p}}{k! (2p-k-1)!} \langle \hat{\phi}^k \rangle \phi_0^{2p-k-1} \right] = 0 ,
\end{split} 
\label{eq-mean-val}
\end{equation}
while, on the other hand, subtracting this last equation to the equation of motion \eqref{EOM}, there remains the equation for the fluctuation $\hat{\phi}$:
\begin{eqnarray}
N\sqrt{g} \Biggl[ \left( \square_L(x) - m^2 \right) \hat{\phi} - \sum_{p=2}^{p_{max}} \sum_{k=0}^{2p-1} \frac{\lambda_{2p}}{k! (2p-k-1)!} \left( \hat{\phi}^k - \langle \hat{\phi}^k \rangle \right) \phi_0^{2p-k-1} \Biggr] = 0 .
\label{eq-fluc}
\end{eqnarray}
In the 1-loop approximation, the contributions of the  expectation values $\langle \hat{\phi}^3 \rangle$, $\langle \hat{\phi}^4 \rangle$, etc., are much smaller than $\langle \hat{\phi}^2 \rangle$. Hence, only the terms with $k=0$ and $k=2$ of the sum in Eq. \eqref{eq-mean-val} and the term with $k=1$ of the sum in Eq. \eqref{eq-fluc} are kept. Then the most important quantum correction to the evolution equation for $\phi_0$ is proportional to $\langle \hat{\phi}^2 \rangle$, 
\begin{eqnarray}
N\sqrt{g} \Biggl[ \left( \square_L(x) - m^2 - \frac{\lambda_{4}}{2} \langle \hat{\phi}^2 \rangle \right) \phi_0 - \sum_{p=2}^{p_{max}} \left( \lambda_{2p} + \frac{\lambda_{2p+2}}{2} \langle \hat{\phi}^2 \rangle \right)  \frac{\phi_0^{2p-1}}{(2p-1)!} \Biggr] \simeq 0 ,
\label{eq-mean-val-1loop}
\end{eqnarray}
while the fluctuation $\hat{\phi}$ satisfies a free field equation with variable mass
\begin{eqnarray}
N\sqrt{g} \left[ \square_L(x) - M^2(x) \right] \hat{\phi}(x) \simeq 0 ,
\label{eq-fluc-1loop} 
\end{eqnarray}
where
\begin{equation}
M^2(x) = m^2 + \sum_{p=2}^{p_{max}} \frac{\lambda_{2p}}{(2p-2)!} \phi_0^{2p-2}(x) .
\label{var-mass}
\end{equation}
In Eq. \eqref{eq-mean-val-1loop}, the sum has been rearranged to group the terms proportional to the same power of $\phi_0$.

\section{Adiabatic expansion  of the quantum corrections}\label{quantum-corr}

To study the renormalization of Eq. \eqref{eq-mean-val-1loop}, it is sufficient to calculate the quantum corrections in an adiabatic expansion, i.e. derivatives of the metric. Each order in this expansion is less divergent than the previous one, so the full ultraviolet behavior of the corrections is exhibited in the first orders of this expansion. In the case of $z=3$, as we already mentioned, the only divergence we can get is logarithmic, so in principle, only the zeroth adiabatic order correction would be divergent so that the renormalization procedure could take place exactly as in flat spacetime. Nevertheless, it is interesting to calculate higher adiabatic order corrections; on the one hand, to find the couplings with the curvature generated by the quantum effects, and on the other hand, to see whether the usual $z=1$ case is recovered when the appropriate limit is taken. In this last case, it is necessary to go up to second adiabatic order when performing the renormalization procedure at the level of the evolution equation \cite{FDM-JPP}. 

The expectation value of $\hat{\phi}^2$ taken in the vacuum state $| 0 \rangle$ can be calculated from the Feynman propagator $G_{\textup{F}}(x,x')$, 
\begin{equation}
\langle 0 | \hat{\phi}^2 (x) | 0 \rangle = -\lim_{x' \to x} \textup{Im}~ G_{\textup{F}}(x,x') ,
\end{equation}
for which it is necessary to solve the equation for the Green function of the fluctuation, which is  
\begin{equation}
N\sqrt{g} \left[ \square_L(x) - M^2(x) \right] G_F(x,x')= \delta(x-x') .
\label{eq-Green-1loop}
\end{equation}

Throughout this paper, the mass $M$ will be taken as constant. An improvement on this approximation would be to consider a derivative expansion on the mass, much like the adiabatic expansion in derivatives of the metric, which in this case, would be equivalent to a derivative expansion on the mean value of the field $\phi_0$. However, the terms including derivatives of $\phi_0$ will not be relevant to analyze the renormalization. 

We will consider, in turn, two different approximations to simplify the calculation of $\langle \hat{\phi}^2 \rangle$ up to second adiabatic order. First (in this section) is a weak gravitational field approximation, and after  that, we will choose an ansatz for the metric and perform a derivative expansion (see Sec. \ref{healthyterms}). In both cases, the goal is to solve for the Green function in the adiabatic expansion.

We first consider small perturbations around flat spacetime:
\begin{equation}
N=1+\delta n ,~~~~~~~~ N^i = \delta N^i ,~~~~~~~ g_{ij} = \delta_{ij} + h_{ij} ,
\end{equation}
keeping terms up to linear order in these perturbations. The Feynman propagator at zero order in the metric perturbations reads
\begin{equation}
G^{(0)}_{\textup{F}}(x,x') = - \int \frac{d^4 k}{(2\pi)^4} ~ \frac{e^{i\, k\cdot(x-x')}}{\left( \omega_k^2 - k_0^2 - i\varepsilon \right)} ,
\label{G0} 
\end{equation}
where
\begin{equation}
\omega_k = \sqrt{M^2 + 2 b_1 |\vec{k}|^2 + 2 \frac{b_2}{\Lambda^2} |\vec{k}|^4 + 2 \frac{b_3}{\Lambda^4} |\vec{k}|^6} .
\label{disp-rel} 
\end{equation}
The first order contribution, following the steps from Ref. \cite{DNL-GG-FDM}, can be written as follows:
\begin{equation}
G_F^{(1)}(x,x') = \int d^4 x'' \frac{d^4 k_1}{(2\pi)^4} \frac{d^4 k_2}{(2\pi)^4} \frac{d^4 p}{(2\pi)^4} ~ \frac{e^{i\, k_1\cdot(x-x'')} \, e^{i\, p\cdot x''} \, e^{i\, k_2\cdot(x''-x')} }{\left( \omega_{k_1}^2 - {k_1}_0^2 \right) \left( \omega_{k_2}^2 - {k_2}_0^2 \right)} ~ f_{k_2}(p) ,
\label{G1-sol} 
\end{equation}
where $f_k(p)$ is a function of $k_0$ , $k_i$ , $p_0$ , and $p_i$ which is linear in the metric perturbations and can be found in the Appendix. To obtain the different orders of the adiabatic expansion, the integrand of Eq. \eqref{G1-sol} must be expanded in powers of $p_i$ and $p_0$. The coincidence limit can be taken trivially and, after Wick rotating to $k_4 = - i k_0$, the integral in $k_4$ can be performed easily. After a straightforward but tedious calculation, the following expression for the zeroth adiabatic order of the coincidence limit of $G^{(1)}_{\textup{F}}(x,x')$ is reached:
\begin{eqnarray}
[G_{\textup{F}}^{(1)}]^{AD(0)} &=& - \int \frac{d^3 k}{(2\pi)^3} ~ \Biggl\{ D^2_2(k) \left( \frac{h}{2}-\delta n \right) - D^0_2(k) \left( \frac{h}{2}+\delta n \right) \omega _{k}^{2} \notag \\
&&~~~~~+ D^0_2(k) h_{ij}k^{i}k^{j} \frac{d\omega _{k}^{2}}{dk^{2}} \Biggr\} , 
\label{GF-adiab-0-coinc}
\end{eqnarray}
where $[\dots]$ denotes the coincidence limit, and 
\begin{eqnarray}
D_b^a(k) &=& \frac{(i)^{a+1}}{(2\pi)}\, \omega_k^{a+1-2b} ~ \int_{-\infty}^{\infty} \frac{dx~x^a}{(1+x^2)^b} \notag \\
&=&\left\{ 
\begin{array}{l l}
  \frac{i\, (-1)^{a/2}}{(2\pi)}\, \omega_k^{a+1-2b} ~ \frac{\Gamma \left( \frac{a+1}{2} \right)\, \Gamma \left( -\frac{a+1-2b}{2} \right)}{\Gamma \left( b \right)}, ~~~~ &a ~~\textup{even},\\
  0, ~~~~ &a ~~\textup{odd}.  
\end{array} \right.  
\label{int-k0-2}  
\end{eqnarray}
The corresponding expressions for the first and second adiabatic orders can be found in the Appendix. 

Finally, after a few more steps regarding the angular integrals, taking the imaginary part and putting everything together, the results for the adiabatic orders of $\langle \hat{\phi}^2 \rangle$ give
\begin{eqnarray}
\langle \hat{\phi}^2 \rangle^{AD(0)} &=& \frac{\tilde{I}_0}{8\pi^2}, \label{results-ad0-1}\\
\langle \hat{\phi}^2 \rangle^{AD(1)} &=& 0, \\
\langle \hat{\phi}^2 \rangle^{AD(2)} &=& \frac{1}{96\pi^2} \Biggl[ I'_0 \left( \partial^2_t h - 2 \partial_t \partial_i \delta N^i \right) + I_0 \left( \partial_i \partial_j h^{ij} - \partial_i \partial^i h \right) \notag \\
&&~~~~~~~~~ - \left(2 I_0 + \frac{5}{6} I_3 - \frac{5}{3} I_2 \right) \partial^2 \delta n \Biggr] ,
\label{results-ad2-1}
\end{eqnarray}
where we have defined the following integrals:
\begin{eqnarray}
\tilde{I}_0 = \int_0^{\infty} du\, \frac{u^{\frac{D}{2}-1}}{\omega_u} ,~~~~~~~~ I_0  = \int_0^{\infty} du\, \frac{u^{\frac{D}{2}-2}}{\omega_u} ,~~~~~~~~ I'_0  = \int_0^{\infty} du\, \frac{u^{\frac{D}{2}-1}}{\omega_u^3} , \notag \\
I_2 = \int_0^{\infty} du\, \frac{u^{\frac{D}{2}}}{\omega_u^3} \frac{d^2\omega _{u}^{2}}{du^2} ,~~~~~~ I_3 = \int_0^{\infty} du\, \frac{u^{\frac{D}{2}+1}}{\omega_u^3} \frac{d^3\omega _{u}^{2}}{du^3} ,
\label{integrals-I} 
\end{eqnarray}
and $u \equiv k^2$. By simply power counting, one can verify that,  provided  $b_3\neq0$, $\tilde{I}_0$ is logarithmically divergent, while the remaining integrals are finite. Therefore, as expected from power counting, only the zeroth adiabatic order correction is divergent (which is the one involving $\tilde{I}_0$), and hence is the only one we need for analyzing the renormalization process (see the next section). We will come back to the result for the second adiabatic order  in Sec.\ref{sec-curvature} when we study the quantum corrections to the couplings with the curvature.

\section{Renormalization and running of the interaction constants}

In the previous section, we obtained the 1-loop quantum corrections to the evolution equation of $\langle \phi \rangle$ up to second adiabatic order. We found that only the  zeroth adiabatic order correction is divergent. Hence, the renormalization procedure only requires counterterms independent of the curvature, which will cause the renormalization of the interaction constants $\lambda_{2p}$ and the field's mass $m$. 

At high energies, the behavior of the running couplings can be obtained from the minimal subtraction renormalization scheme. The divergent integral $\tilde{I}_0$ is regularized to give a finite expression using dimensional regularization, and the counterterms are choosen to precisely cancel the poles in $D=3$. This process requires the introduction of an arbitrary mass scale $\mu$ in order to keep the physical couplings with the correct dimensions,
\begin{equation}
\lambda_{2p}^B = \mu^{(p-1)\epsilon} (\lambda_{2p}^R + \delta \lambda_{2p}),
\end{equation}
where the superscripts $B$ and $R$ stand for bare and renormalized, respectively, $\delta \lambda_{2p}$ are the counterterms and $\epsilon = 3-D$. Replacing the bare quantities for the renormalized ones in the evolution equation, it reads
\begin{equation}
N\sqrt{g} \Biggl[ \left( \square_L(x) - m_R^2 - \delta m^2 - \frac{\lambda_{4}^R}{2} \mu^{\epsilon} \langle \hat{\phi}^2 \rangle \right) \phi_0 - \sum_{p=2}^{p_{max}} \mu^{(p-1)\epsilon} \left( \lambda_{2p}^R + \delta \lambda_{2p} + \frac{\lambda_{2p+2}^R}{2} \mu^{\epsilon} \langle \hat{\phi}^2 \rangle \right) \frac{\phi_0^{2p-1}}{(2p-1)!} \Biggr] = 0. 
\label{eq-mean-val-renorm-1}
\end{equation}

The evaluation of $\tilde{I}_0$ is more complex than for the usual case, so we sketch here the main steps to extract the divergent part. We start with the integral representation:
\begin{equation}
\tilde I_0=\frac{1}{\sqrt{\pi}} \int_0^\infty ds\int_0^\infty du\, \frac{u^{\frac{D}{2}-1}}{\sqrt{s}} e^{-s\tilde\omega_u^2}\, 
e^{-s\Delta\omega_u^2},\end{equation}
where $\tilde\omega_u^2=M^2+2b_3u^3/\Lambda^4$, $\Delta\omega_u^2=2b_1u+2b_2u^2/\Lambda^2$. Expanding the exponential in powers of $\Delta\omega_u^2$, it is easy to see that only the leading term is divergent for $D\to 3$ and can be explicitly evaluated. Then, performing a series expansion on $\epsilon$, we can write
\begin{equation}
\mu^{\epsilon} \langle \hat{\phi}^2 \rangle = \frac{\Lambda^2}{4 \pi^2 \sqrt{2 b_3}} \left[ \frac{1}{\epsilon} + \ln(\mu) \right] + \textup{finite} + \mathcal{O}(\epsilon),
\end{equation}
where the finite terms are independent of both $\mu$ and $\epsilon$. Inserting this in Eq. \eqref{eq-mean-val-renorm-1}, we choose the counterterms as
\begin{eqnarray}
\delta m^2 &=& - \frac{ \lambda_{4}^R \Lambda^2}{8 \pi^2 \sqrt{2 b_3}}~ \frac{1}{\epsilon} ,\label{mass-counterterm}\\
\delta \lambda_{2p} &=& - \frac{ \lambda_{2p+2}^R \Lambda^2}{8 \pi^2 \sqrt{2 b_3}}~ \frac{1}{\epsilon} ,\label{lambda-counterterm}
\end{eqnarray}
so to cancel only the poles in $\epsilon=0$. Afterwards, the $\epsilon \to 0$ limit can be safely taken, and the renormalized evolution equation reads
\begin{equation}
N\sqrt{g} \Biggl[ \left( \square_L(x) - m_R^2 - \frac{ \lambda_{4}^R \Lambda^2}{8 \pi^2 \sqrt{2 b_3}} \ln{(\mu)} - \textup{finite} \right) \phi_0 - \sum_{p=2}^{p_{max}} \left( \lambda_{2p}^R + \frac{ \lambda_{2p+2}^R  \Lambda^2}{8 \pi^2 \sqrt{2 b_3}} \ln{(\mu)} + \textup{finite} \right) \frac{\phi_0^{2p-1}}{(2p-1)!} \Biggr] = 0. 
\label{eq-mean-val-renorm-2}
\end{equation}
This equation inherits a $\mu$ dependence from the renormalization (or regularization) procedure, which remains even after taking the $\epsilon \to 0$ limit. However, the physics must be independent of the scale so introduced. Therefore, the $\ln(\mu)$ derivative of the previous equation must vanish term-by-term in the sum of powers of $\phi_0$. Then, it follows that renormalized couplings must satisfy the following UV RG equations:
\begin{eqnarray}
\mu \, \frac{d m_R^2}{d \mu} &=& -\frac{\lambda_4^R \Lambda^2}{8 \pi^2 \sqrt{2 b_3}} \label{mR_UV} ,\\ 
\mu \, \frac{d \lambda_{2p}^R}{d \mu} &=& -\frac{\lambda_{2p+2}^R \Lambda^2}{8 \pi^2 \sqrt{2 b_3}} .\label{lambdaR_UV}
\end{eqnarray}

To conclude this section, we introduce the notation $\langle \hat{\phi}^2 \rangle_{ren} = \langle \hat{\phi}^2 \rangle - \langle \hat{\phi}^2 \rangle^{AD(0)}_{div}$ for the finite part of  $\langle \hat{\phi}^2 \rangle^{AD(0)}$ and the contributions of higher adiabatic order. Then, the renormalized evolution equation is 
\begin{eqnarray}
N\sqrt{g} \Biggl[ \left( \square_L(x) - m_R^2 - \frac{\lambda_{4}^R}{2} \langle \hat{\phi}^2 \rangle_{ren} \right) \phi_0 - \sum_{p=2}^{p_{max}} \left( \lambda_{2p}^R + \frac{\lambda_{2p+2}^R}{2} \langle \hat{\phi}^2 \rangle_{ren} \right) \frac{\phi_0^{2p-1}}{(2p-1)!} \Biggr] = 0 .
\label{eq-mean-val-renorm}
\end{eqnarray}
This expression will be picked up later in Sec. \ref{sec-curvature} to discuss the type of couplings with the curvature that come from the second-adiabiatic-order part of $\langle \hat{\phi}^2 \rangle_{ren}$.

\section{Infrared regime}

The minimal subtraction scheme cannot be used to study the infrared behavior of the running couplings, but a more physical renormalization scheme must be used instead, such as the momentum subtraction scheme. Another possibility to study this problem is to use the so-called RG-flow equation for the effective action \cite{erge}. As we will see, this approach is useful not only for obtaining the RG equations for the coupling constants of the potential in the UV and IR regimes, but also to extend the analysis beyond the one-loop approximation.
  
We will therefore generalize here the methods presented in Ref. \cite{Liao-Polonyi} to the Lifshitz case. In this formalism, the 1-loop correction to the effective potential can be written as
\begin{eqnarray} U_\mu^{(1)}(\Phi) &=& \frac{1}{2\Omega} \ln \Biggl\{ \det \left[ \frac{\partial^2 S_\mu}{\partial \phi(x) \partial \phi(y)} \Bigg|_{\Phi} \right] \Biggr\} \\
&=& \frac{1}{2} \int_{\mu<|k|<\Lambda_{UV}} \frac{d^4 k}{(2\pi)^4}~\Biggl\{ \ln \left[ k^2 + V^{''}(\Phi) \right] - \ln (k^2) \Biggl\} \notag \\
&=& \frac{1}{16 \pi^2} \int_\mu^{\Lambda_{UV}} dk~k^3~\ln \left[ 1 + \frac{V^{''}(\Phi)}{k^2} \right] ,
\label{U-usual}
\end{eqnarray}
where the integral is performed over the Euclidean momentum space $k$. The renormalization scale $\mu$ is identified with the IR momentum cutoff, so that the integration region is the volume contained between two concentric 3-spheres of radii $\mu$ and $\Lambda_{UV}$. 

The main difference with the Lifshitz case is the anisotropic scaling. This is evidenced in the integrand, which, with the modified dispertion relation \eqref{disp-rel}, gets the replacement 
\begin{equation}
k^2 \to -k_0^2 + \tilde{\omega}^2(\vec{k}) = k_4^2 + \tilde{\omega}^2(\vec{k}) ,
\end{equation}
with $\tilde{\omega}^2(\vec{k}) = \omega^2(\vec{k}) - M^2$ [the effective mass term $M^2$ is included in the potential $V(\Phi)$], and then it is no longer spherically symmetric in 4 dimensions. Hence, the anisotropic scaling should be taken into account by modifying also the integration region, using different cutoffs in the spatial and temporal directions.

We motivate one way of performing this generalization by first looking at the integral over the 3-sphere from the usual Lorentz-invariant case. This integral can be written
\begin{eqnarray}
 \int_\mu^{\Lambda_{UV}} dk~k^3 = \int_0^{\Lambda_{UV}} dk~k^3 - \int_0^{\mu} dk~k^3 &=& \int_0^{\Lambda_{UV}} dk~|\vec{k}|^2 \int_{-\sqrt{\Lambda_{UV}^2-|\vec{k}|^2}}^{\sqrt{\Lambda_{UV}^2-|\vec{k}|^2}} dk_4 \notag \\ &&- \int_0^{\mu} dk~|\vec{k}|^2 \int_{-\sqrt{\mu^2-|\vec{k}|^2}}^{\sqrt{\mu^2-|\vec{k}|^2}} dk_4 .
\end{eqnarray}
In this form, we can introduce the anisotropy by modifying the limits of the $k_4$ integral in the following manner:
\begin{equation}
\sqrt{\Lambda_{UV}^2-|\vec{k}|^2} \to \sqrt{\tilde{\omega}^2(\Lambda_{UV})-\tilde{\omega}^2(\vec{k})} ~~~~~~~~ ; ~~~~~~~~ \sqrt{\mu^2-|\vec{k}|^2} \to \sqrt{\tilde{\omega}^2(\mu)-\tilde{\omega}^2(\vec{k})} .
\end{equation}
This leads to an integral over the volume contained between two sort of prolate 3-spheroids, with their major axes aligned in the $k_4$ direction, and whose cross sections in the plane perpendicular to $k_4$ are concentric 2-spheres. This 3-``rugby-ball'' can be continuously deformed into a 3-sphere when $\tilde{\omega}^2(\vec{k}) \to |\vec{k}|^2$, which is expected to happen in the IR.

Putting all this together, and taking into account that the integrand now only depends on $|\vec{k}|$ and $k_4^2$, the expression \eqref{U-usual} is generalized to
\begin{eqnarray}
U_\mu^{(1)}(\Phi) &=& \frac{1}{4 \pi^3} \int_0^{\Lambda_{UV}} dk~|\vec{k}|^2~\int_0^{\sqrt{\tilde{\omega}^2(\Lambda_{UV})-\tilde{\omega}^2(\vec{k})}} dk_4~\ln \left[ 1 + \frac{V^{''}(\Phi)}{k_4^2 + \tilde{\omega}^2(\vec{k})} \right] \notag \\ &&- \frac{1}{4 \pi^3} \int_0^{\mu} dk~|\vec{k}|^2~\int_0^{\sqrt{\tilde{\omega}^2(\mu)-\tilde{\omega}^2(\vec{k})}} dk_4~ \ln \left[ 1 + \frac{V^{''}(\Phi)}{k_4^2 + \tilde{\omega}^2(\vec{k})} \right] .
\label{U-Lifshitz}
\end{eqnarray}
We cannot go much further from here to obtain $U_{\mu}^{(1)}(\Phi)$ because of calculational difficulties. Nevertheless, in order to calculate  the RG equations,  we can take the $\ln(\mu)$ derivative of $U_{\mu}^{(1)}(\Phi)$ before performing the integrals, which greatly simplifies the computations. 

The renormalized constants are obtained by taking a given number of $\Phi$ derivatives of $U_{\mu}(\Phi)$ and then evaluating at $\Phi=0$:
\begin{eqnarray}
m_R^2(\mu) &=& \frac{\partial^2 U_{\mu}}{\partial \Phi^2} \bigg|_{\Phi=0} ,\\ 
\lambda_4^R(\mu) &=& \frac{\partial^4 U_{\mu}}{\partial \Phi^4} \bigg|_{\Phi=0} ,
\end{eqnarray}
and, in general,
\begin{equation}
\lambda_{2p}^R(\mu) = \frac{\partial^{2p} U_{\mu}}{\partial \Phi^{2p}} \bigg|_{\Phi=0} .
\end{equation}
From the expression \eqref{U-Lifshitz}, the first term is independent of $\mu$ and is dropped, while in the second, there are two nested integrals whose limits depend on $\mu$. Performing the derivatives, we get
\begin{equation}
\mu ~ \frac{\partial U_\mu(\Phi)}{\partial \mu} =- \frac{\mu}{4 \pi^3}  \frac{d \tilde{\omega}^2}{dk^2}\Bigg|_{k=\mu} \Biggl\{ \int_0^{\mu} dk~|\vec{k}|^2~ \ln \left[ 1 + \frac{V^{''}(\Phi)}{k_4^2 + \tilde{\omega}^2(\vec{k})} \right]~ \frac{\mu}{\sqrt{\tilde{\omega}^2(\mu)-\tilde{\omega}^2(\vec{k})}} \Biggr\}_{k_4 = \sqrt{\tilde{\omega}^2(\mu)-\tilde{\omega}^2(\vec{k})}} .
\end{equation}
On the right-hand side, the logarithm becomes independent of $|\vec{k}|$ once we evaluate $k_4 = \sqrt{\tilde{\omega}^2(\mu)-\tilde{\omega}^2(\vec{k})}$. Hence, we can take it out of the integral.  What remains is quite simple:
\begin{eqnarray}
\mu ~ \frac{\partial U_\mu(\Phi)}{\partial \mu}  = - f(\mu,\{b_i\}) ~ \ln \left[ 1 + \frac{V^{''}(\Phi)}{\tilde{\omega}^2(\mu)} \right] ,
\label{dkU-1loop}
\end{eqnarray}
where we have defined
\begin{equation}
f(\mu,\{b_i\}) \equiv \frac{\mu^2}{4 \pi^3}  \frac{d \tilde{\omega}^2(\mu)}{d\mu^2} \int_0^{\mu} dk~ \frac{|\vec{k}|^2}{\sqrt{\tilde{\omega}^2(\mu)-\tilde{\omega}^2(\vec{k})}}.
\label{integral-mu}
\end{equation}
For the dispertion relation $\tilde{\omega}^2(k) = 2 b_1 k^2 + 2 b_2 k^4/\Lambda^2 + 2 b_3 k^6/\Lambda^4 $, this function $f(\mu,\{b_i\})$ cannot be calculated analytically, making it difficult to reach expressions valid for any value of $\mu$. However, it is possible to study its asymptotic behavior in the UV and IR limits. On the one hand, in the UV we have $\mu \gg \Lambda$ and therefore $\tilde{\omega}^2(k) \simeq \frac{2 b_3}{\Lambda^4} k^6$, which leads to
\begin{equation}
f_{UV} \simeq f(\mu,b_3) = \frac{\sqrt{2 b_3}}{8 \pi^2} \frac{\mu^6}{\Lambda^2}, 
\label{fmu-UV}
\end{equation}
while on the other hand, in the IR,  $\mu \ll \Lambda$ so that $\tilde{\omega}^2(k) = 2 b_1 k^2$, giving
\begin{equation}
f_{IR} \simeq f(\mu,b_1) = \frac{\sqrt{2 b_1}}{16 \pi^2} \mu^4.
\label{fmu_IR}
\end{equation}

Up to here, we have kept the 1-loop corrections to the effective potential. A RG improvement \cite{Liao-Polonyi} can be made on Eq. \eqref{dkU-1loop} by replacing $V^{''}(\Phi)$ by $\partial^2_{\Phi} U_{\mu}$ on the right-hand side, leading to
\begin{eqnarray}
\mu ~ \frac{\partial U_\mu(\Phi)}{\partial \mu}  = - f(\mu,\{b_i\}) ~ \ln \left[ 1 + \frac{\partial^2_{\Phi} U_{\mu}}{\tilde{\omega}^2(\mu)} \right].
\label{dkU}
\end{eqnarray}
When taking the $\Phi$ derivatives and evaluating at $\Phi = 0$ to find the RG equations for $m_R^2$, $\lambda_4^R$ or any $\lambda_{2p}^R$, only the logarithm in the previous expression comes into play. Within the resulting expressions, there will be terms containing different number of $\Phi$ derivatives of $U_{\mu}$, which might vanish upon evaluation at $\Phi = 0$. For instance, for the potential considered here \eqref{action}, odd derivatives of $U_{\mu}$ will evaluate to zero, as also will any number of derivatives greater than $n_{max}$.

We start by considering the RG equation for the renormalized mass $m^2_R$. After the appropriate derivation and evaluation at $\Phi=0$ of Eq. \eqref{dkU}, it reads
\begin{equation}
\mu ~ \frac{d m^2_R}{d \mu} = - f(\mu,\{b_i\}) ~ \frac{\lambda_4^R}{\left( m^2_R + \tilde{\omega}^2(\mu) \right)}.
\end{equation}
The asymptotic behavior of this equation is obtained by replacing $\tilde{\omega}^2(\mu)$ and $f(\mu,\{b_i\})$ by their corresponding expressions in the UV or in the IR. The results are
\begin{eqnarray}
\mu ~ \frac{d m_R^2}{d \mu} &=& -\frac{\lambda_4^R \Lambda^2}{8 \pi^2 \sqrt{2 b_3}} , ~~~~~~~~~~~~~~~~~~~~~ \Lambda \ll \mu ~~ (\textup{UV}), \\
\mu ~ \frac{d m_R^2}{d \mu} &=& -\frac{\sqrt{2 b_1}}{16 \pi^2} \frac{\lambda_4^R \, \mu^4}{\left( m^2_R + 2 b_1 \mu^2 \right)} , ~~~~~~~ \mu \ll \Lambda ~~ (\textup{IR}).
\end{eqnarray}
While the first equation is the same that was obtained earlier by dimensional regularization and minimal substraction \eqref{mR_UV}, the second equation reproduces the standard result for the Lorentz-invariant case \cite{Liao-Polonyi} for $b_1 \to 1/2$.

In a similar fashion, we obtain the RG equation for $\lambda_4^R$ by taking four derivatives prior to the evaluation at $\Phi = 0$. The result is
\begin{equation}
\mu ~ \frac{d \lambda_4^R}{d \mu} = - f(\mu,\{b_i\}) ~ \left[ \frac{\lambda_6^R}{\left( m^2_R + \tilde{\omega}^2(\mu) \right)} - \frac{3 (\lambda_4^R)^2}{\left( m^2_R + \tilde{\omega}^2(\mu) \right)^2} \right],
\end{equation}
and the corresponding UV and IR asymptotic expressions are
\begin{eqnarray}
\mu ~ \frac{d \lambda_{4}^R}{d \mu} &=& -\frac{\lambda_{6}^R \Lambda^2}{8 \pi^2 \sqrt{2 b_3}} + \frac{3 (\lambda_{4}^R)^2}{8 \pi^2 (2b_3)^{3/2}} \left(\frac{\Lambda}{\mu} \right)^6, ~~~~~~~~~~~~~~~~~~~~~~~~ \Lambda \ll \mu ~~ (\textup{UV}), \\
\mu ~ \frac{d \lambda_{4}^R}{d \mu} &=& -\frac{\sqrt{2 b_1}}{16 \pi^2} \frac{ \lambda_{6}^R \, \mu^4}{\left( m^2_R + 2 b_1 \mu^2 \right)} + \frac{3 \sqrt{2 b_1}}{16 \pi^2} \frac{(\lambda_{4}^R)^2  \, \mu^4}{\left( m^2_R + 2 b_1 \mu^2 \right)^2}, ~~~~~~~ \mu \ll \Lambda ~~ (\textup{IR}).
\end{eqnarray}
As the second term in the right-hand side of the UV equation can be neglected in this limit, it again coincides with the one found previously \eqref{lambdaR_UV} for $p=2$, while for the IR equation, we recover the usual case provided $b_1 \to 1/2$.

So far, we have dealt with the mass and a coupling that are both renormalizable in the usual case. We would like to see what happens with the new couplings that are not renormalizable in that case. However, for couplings with $p>2$, the corresponding expressions get more and more complicated as a higher number of $\Phi$ derivatives are taken, generating increasingly more terms. This can be thought diagramatically as the increasing number of ways to obtain a $2p$-legged diagram at 1 loop for higher $p$'s, because there are more types of vertices with fewer legs available. 

Nevertheless, let us consider $\lambda_6^R$ as the easiest example of a coupling that it is not renormalizable in the usual case. Its RG equation is
\begin{equation}
\mu ~ \frac{d \lambda_6^R}{d \mu} = - f(\mu,\{b_i\}) \left[ \frac{\lambda_8^R}{\left( m^2_R + \tilde{\omega}^2(\mu) \right)} - \frac{15 \lambda_4^R \lambda_6^R}{\left( m^2_R + \tilde{\omega}^2(\mu) \right)^2} + \frac{30 (\lambda_4^R)^3 }{\left( m^2_R + \tilde{\omega}^2(\mu) \right)^3} \right],
\end{equation}
and the UV and IR limits are, respectively,
\begin{eqnarray}
\mu ~ \frac{d \lambda_{6}^R}{d \mu} &=& -\frac{\lambda_{8}^R \Lambda^2}{8 \pi^2 \sqrt{2 b_3}} + \frac{15 \lambda_{4}^R \lambda_{6}^R}{8 \pi^2 (2b_3)^{3/2}} \left(\frac{\Lambda}{\mu} \right)^6 - \frac{30 (\lambda_{4}^R)^3}{8 \pi^2 (2b_3)^{5/2} \mu^2} \left(\frac{\Lambda}{\mu} \right)^{10}, ~~~~~~~~~ \Lambda \ll \mu ~~ (\textup{UV}), \\
\mu ~ \frac{d \lambda_{6}^R}{d \mu} &=& -\frac{\sqrt{2 b_1}}{16 \pi^2} \frac{ \lambda_{8}^R \, \mu^4}{\left( m^2_R + 2 b_1 \mu^2 \right)} + \frac{15 \sqrt{2 b_1}}{16 \pi^2} \frac{\lambda_{4}^R \lambda_{6}^R  \, \mu^4}{\left( m^2_R + 2 b_1 \mu^2 \right)^2} \nonumber\\
&& \,
- \frac{30 \sqrt{2 b_1}}{16 \pi^2} \frac{(\lambda_{4}^R)^3 \, \mu^4}{\left( m^2_R + 2 b_1 \mu^2 \right)^3}, ~~~~~~~~~~~~~~~~~~~~~~~~~~~~~~~~~~~~~~~~~~~~~~~~~~~~ \mu \ll \Lambda ~~ (\textup{IR}).
\label{l6IR}
\end{eqnarray}
Once more in the UV equation, the second and third terms in the right-hand side can be dropped in this limit, and we get the same as with dimensional regularization and minimal substraction, that is Eq.\eqref{lambdaR_UV} for $p=3$. In the IR limit, the equation is the usually encountered in effective field theories at low energies, again for $b_1 \to 1/2$.

Higher order couplings get more and more complicated expressions. A full analysis of the running of the coupling constants would involve a solution of the coupled differential equations for all couplings.

\section{Couplings to the curvature}\label{sec-curvature}

We will now consider the coupling of the scalar field to terms containing second derivatives of the metric. In the $z=1$ case, $\langle\hat\phi^2\rangle$ contains a divergence proportional to the Ricci scalar, and therefore the corresponding counterterm $\xi {}^{(4)}R \phi^2 $ should be introduced in the action. For $z=3$, the second adiabatic order correction \eqref{results-ad2-1} is finite. These terms can be grouped to form the linearized expressions for the following curvature invariants:
\begin{eqnarray}
^{(3)}R &\simeq& \partial_i \partial_j h^{ij} -\partial_i \partial^i\, h , \label{ricci3-linear} \\
K &\simeq& \frac{1}{2} \left( \dot{h} - 2 \partial_i \delta N^i \right) ,
\end{eqnarray}
where $^{(3)}R$ is the scalar of 3-curvature, and $K$ is the trace of the extrinsic curvature $K_{ij}$. The dot stands for time derivative. With these identifications and using the following relation valid at linear order on the metric perturbations
\begin{equation}
^{(4)}R \simeq \, ^{(3)}R + 2 \left( \dot{K} - \partial_i \partial^i \delta n \right) ,
\label{ricci4-descomp-linear} 
\end{equation}
we can rewrite Eq. \eqref{results-ad2-1} as
\begin{eqnarray}
\langle \hat{\phi}^2 \rangle^{AD(2)} &=& A~ ^{(4)}R + B \, \dot{K} + C \left( \frac{5}{3} \partial^2 \delta n - 2 \dot{K} \right) .
\label{results-ad2-3}
\end{eqnarray}
Here, the coefficients $A$, $B$ and $C$ come from the integrals defined in Eq. \eqref{integrals-I}, and their expressions are
\begin{eqnarray}
A &=& \frac{1}{96\pi^2} \int_0^{\infty} dx\, \frac{x^{-1/2}}{\left[M^2/\Lambda^2 + 2b_1 x + 2b_2 x^2 + 2b_3 x^3\right]^{1/2}} , \label{coeff-A}\\ 
B &=& \frac{(1-2b_1)}{48\pi^2} \int_0^{\infty} dx\, \frac{x^{1/2}}{\left[M^2/\Lambda^2 + 2b_1 x + 2b_2 x^2 + 2b_3 x^3\right]^{3/2}} ,\\ 
C &=& \frac{1}{96\pi^2} \int_0^{\infty} dx\, \frac{x^{3/2} (4b_2 + 6b_3 x)}{\left[M^2/\Lambda^2 + 2b_1 x + 2b_2 x^2 + 2b_3 x^3\right]^{3/2}} .
\end{eqnarray}
The renormalized evolution equation \eqref{eq-mean-val-renorm} then reads
\begin{eqnarray}
N\sqrt{g} \Biggl\{&& \left( \square_L(x) - m_R^2 - \frac{\lambda_4^R}{2} \langle \hat{\phi}^2 \rangle_{ren}^{AD(0)} \right) \phi_0  - \sum_{p=2}^{p_{max}} \left( \lambda_{2p}^R + \frac{\lambda_{2p+2}^R}{2} \langle \hat{\phi}^2 \rangle_{ren}^{AD(0)} \right) \frac{\phi_0^{2p-1}}{(2p-1)!} \nonumber\\
&&~~~~~~- \sum_{p=1}^{p_{max}} \frac{\lambda_{2p+2}^R}{2} \Biggl[ A~ ^{(4)}R + B \, \dot{K} + C \left( \frac{5}{3} \partial^2 \delta n - 2 \dot{K} \right) \Biggr] \frac{\phi_0^{2p-1}}{(2p-1)!} \Biggr\} \simeq 0 .
\end{eqnarray}

Had we taken the parameters $b_1 = 1/2$, $b_2 = b_3 = 0$, or $\Lambda\to\infty$ from the beginning, the coefficients $B$ and $C$ would be identically zero, and the correction \eqref{results-ad2-3} would have been equal to the standard result for the Lorentz-invariant  theory (see Ref. \cite{FDM-JPP}), that is
\begin{eqnarray}
\langle \hat{\phi}^2 \rangle^{AD(2)} = A_{usual}~ ^{(4)}R ,
\end{eqnarray}
where $A_{usual}$ is $A$ with its integrand evaluated for these values of the parameters before integration, and it is a divergent integral. However, if we consider a Lorentz-noninvariant UV completion of the theory like the one we are studying in this work, the situation is different. When taking the infrared limit, which is $\Lambda \to \infty$, the coefficients $A$, $B$ and $C$ do not vanish, not even in the particular case $b_1=1/2$, $b_2, b_3\to 0$.

On the one hand, the coefficient $B$ can take various values depending on $b_1$'s initial value. If $b_1$ is precisely $1/2$, then $B=0$, but for any other case, it will depend on the particular running of $b_1$, $b_2$ and $b_3$, with $B \to \infty$ being a possibility. To study this,  the running of $b_1$ should be studied in more detail.

On the other hand, the coefficient $C$ is given by an integral that is finite regardless of the mass $M$, so it can be calculated explicitly by taking the $\Lambda \to \infty$ limit (or $M \to 0$) prior to integration. The result is finite,
\begin{equation}
C = \frac{\sqrt{2} (2 b_2 + 3 \sqrt{b_1 b_3})}{\sqrt{b_1} (b_2 + 2 \sqrt{b_1 b_3})} ,
\end{equation}
and has nonvanishing limits even when $b_2 \to 0$ and $b_3 \to 0$, effectively leaving a non generally-covariant remnant in the 1-loop evolution equation of $\langle \phi \rangle$ that does not vanish in the IR, in spite of the parameters taking their Lorentz-invariant values. Similar effects have been observed before, for example, in the case of a theory with a Yukawa-type scalar-fermion interaction in flat spacetime and with a generically modified dispersion relation for the fermion at high energies \cite{vucetich}.

We will now take a closer look to the integral \eqref{coeff-A}, that defines the $A$ coefficient to try to find out its behavior when $\Lambda \to \infty$. In this limit, the integral has an infrared divergence; however, away from it, $\kappa \equiv M/\Lambda$ acts as an infrared cutoff. By noting that the integrand is sensitive to the value of $\kappa \equiv M/\Lambda$ only when $x \ll 1$, we can separate the integration interval in two parts, $0 \leq x \leq \epsilon \ll 1$ and $\epsilon \leq x < \infty$. In the first region, we can drop $x^3$ and $x^4$ in front of $x^2$, while in the second region, we no longer have the infrared divergence and, consequently, we can safely take the limit $\kappa \to 0$. Then the coefficient $A$ approximately reads
\begin{equation}
A \simeq \frac{1}{96\pi^2} \Biggl[ \int_0^{\epsilon} dx\, \frac{1}{\sqrt{\kappa^2 x + 2b_1 x^2}} + \int_{\epsilon}^{\infty} dx\, \frac{1}{\sqrt{2b_1 x^2 + 2b_2 x^3 + 2b_3 x^4}}\Biggr], \label{coeff-A-2}\\ 
\end{equation}
which now contains two integrals that can be performed analytically, giving
\begin{equation}
A \simeq \frac{1}{96\pi^2 \sqrt{b_1}} \Biggl[ \frac{1}{\sqrt{2}} \textup{ArcCosh}{\left( 1 + \frac{4 b_1 \epsilon}{\kappa^2} \right)}  + \ln{\left( \frac{2 b_1 + b_2 \epsilon + 2 \sqrt{ b_1 (b_1 + b_2 \epsilon + b_3 \epsilon^2)} }{b_2 \epsilon + 2\sqrt{b_1 b_3} \epsilon} \right)} \Biggr]. \label{coeff-A-3}\\ 
\end{equation}
The dependence in $\kappa$, and ultimately on $\Lambda$, is contained in the first part. As we are interested in the behavior for $\kappa \to 0$, we perform a series expansion in $\kappa$ around zero. Meanwhile, the second part is finite and of order 1 for any positive sufficiently small nonvanishing value of $\epsilon$, which is when this treatment gives a closer estimate of $A$. Finally, we can say that the behavior of this coefficient for large $\Lambda$ is approximately
\begin{equation}
A \simeq \frac{1}{96\pi^2 \sqrt{b_1}} \Biggl[ \sqrt{2} \ln{\left( \frac{\Lambda}{M} \right)} + \mathcal{O}(1) \Biggr], \label{coeff-A-4}\\ 
\end{equation}
which shows that the constant $A$, divergent in the Lorentz-invariant theory, is now finite with $\Lambda$ acting as a cutoff. The dependence is logarithmic, which means that the corrections are not large unless $M$ is many orders of magnitude below $\Lambda$.

\subsection{Coupling to ``healthy" terms}\label{healthyterms}

According to Ref. \cite{blas}, the Ho\v{r}ava-Lifshitz theory of gravity is better behaved when invariants formed with $a_i = \partial_i \ln(N)$ are included in the gravitational action than when they are not. The most relevant invariant is $a_i a^i$, which is absent in general relativity. 

It is interesting to see whether the self-interaction of the scalar field induces couplings to such terms or not. In order to do that, it is necessary to go beyond the weak field approximation considered so far. Instead of tackling this complicated calculation, we will follow a shortcut by performing a derivative expansion on $N$, taking as a starting point a metric of the form
\begin{equation}
ds^2 = - N^2(\vec{x}) dt^2 + d\vec{x}^2,
\label{metric-NofX} 
\end{equation} which is enough for our present purpose.
The corresponding equation for the Green function is
\begin{equation}
\begin{split}
- \frac{1}{N} \partial^2_t G(x,x') - N M^2 G(x,x') + 2 b_1 \partial_i \left( N \partial^i G(x,x') \right)&\\ - \frac{2 b_2}{\Lambda^2} \partial^2 \left( N \partial^2 G(x,x') \right) + \frac{b_3}{\Lambda^4} \partial^2 \left( N \partial^4 G(x,x') \right) + \frac{b_3}{\Lambda^4} \partial^4 \left( N \partial^2 G(x,x') \right)& = \delta(x-x') .
\end{split} 
\label{ec-func-Green-ansatz} 
\end{equation}
After expanding in spatial derivatives of $N$ up to second order and then going to Fourier space, the equation reads
\begin{eqnarray}
\left( \frac{k_0^2}{N^2} - \omega_k^2 \right) \tilde{G}(k) -i \left[ \left( \frac{k_0^2}{N^2} + \omega_k^2 \right) \frac{\partial \tilde{G}(k)}{\partial k_i} + \frac{d \omega_k^2}{d k^2} k^i \tilde{G}(k) \right] a_i && \notag \\
+ \frac{1}{2} \Biggl[ \left( \frac{k_0^2}{N^2} + \omega_k^2 \right) \frac{\partial^2 \tilde{G}(k)}{\partial k_i \partial k_j} + \frac{d \omega_k^2}{d k^2} \left( k^i \frac{\partial \tilde{G}(k)}{\partial k_j} + k^j \frac{\partial \tilde{G}(k)}{\partial k_i} \right) && \label{ec-green-Fourier-ansatz} \\
+ \frac{d^2 \omega_k^2}{d (k^2)^2} k^2 \delta^{ij} \tilde{G}(k) + \frac{d^3 \omega_k^2}{d (k^2)^3} \left( \frac{2}{3} k^i k^j - \frac{k^2}{2} \delta^{ij} \right) k^2 \tilde{G}(k) \Biggr] b_{ij} - \frac{k_0^2}{N^2} \frac{\partial^2 \tilde{G}(k)}{\partial k_i \partial k_j} a_i a_j &=& \frac{1}{N} . \notag
\end{eqnarray}
To solve it, the solution is written as a sum of contributions of ascending order in derivatives of $N$, noted with a subscript $Ni$, so
\begin{equation}
\tilde{G}(k) = \tilde{G}_{N0}(k) + \tilde{G}_{N1}(k) + \tilde{G}_{N2}(k) + \dots ,
\label{G-orders-N}
\end{equation}
and then the equation is solved iteratively order by order. The expressions obtained for $\tilde{G}_{N0}(k)$, $\tilde{G}_{N1}(k)$ and $\tilde{G}_{N2}(k)$ are given in the Appendix. 

The calculation continues from this point in the same manner as in the weak gravitational field approximation, giving the following results:
\begin{eqnarray}
\langle \hat{\phi}^2 \rangle_{N0} &=& \frac{\tilde{I}_0}{8\pi^2} ,\\
\langle \hat{\phi}^2 \rangle_{N1} &=& 0 ,\\
\langle \hat{\phi}^2 \rangle_{N2} &=& - \frac{1}{96\pi^2} \Biggl[ 2 I_0 + \frac{5}{6} I_3 - \frac{5}{3} I_2 \Biggr] \frac{\partial^2 N}{N} ,
\label{results-N2-1}
\end{eqnarray}
where the integrals $\tilde{I}_0$, $I_0$, $I_2$ and $I_3$ are the same as defined earlier in Eq. \eqref{integrals-I}. Notably, the coefficient accompanying $a_i a^i$ has vanished, so the result is just the same as the one obtained in the previous section \eqref{results-ad2-1}.

Evidently, the absence of this kind of correction is more fundamental than a simple over-approximation. In fact, it can be easily verified that the result would be the same if we had taken the Lorentz-breaking terms in the action to be slightly different, as we will now see.

\subsection{More general Lagrangians}\label{moregenLag}

We already found in the previous section that couplings between the scalar field and $a_i a^i$  seem not to be generated. However, one can ask if this is a consequence of the particular form of the action we started with Eq. \eqref{action}. Indeed, we could have chosen any term of the form
\begin{equation}
N \sqrt{g} \partial^a \phi \partial^b \phi ,
\end{equation}
where  the superscripts $a$ and $b$ denote the number  of spatial derivatives acting on the field which satisfy $a+b = 2n \leq z$, regardless of how the indexes $i,j..$ of each pair of derivatives are contracted. Performing a variation of the action, this kind of term leads to
\begin{eqnarray}
\frac{\delta}{\delta \phi} \left( \int d^3 x dt \, N \partial^a \phi\, \partial^b \phi \right) &=& \int d^3 x dt \, \left[ (-1)^a \partial^a (N \partial^b \phi) + (-1)^b \partial^b (N \partial^a \phi) \right] \notag \\
&=& (-1)^a \int d^3 x dt \, \left[ 2N \partial^{a+b} \phi + (a+b) \partial_i N \partial^i \partial^{a+b-1} \phi + \tilde{L}[\Delta N, \phi] \right] \notag \\
&=& (-1)^a \int d^3 x dt \, \left[ 2N \partial^{2n} \phi + 2n\, \partial_i N \partial^i \partial^{2n-1} \phi + \tilde{L}[\Delta N, \phi] \right] , 
\end{eqnarray}
where  $\tilde{L}$ contains terms with second-order-or-higher spatial derivatives of $N$, and its form will be dependent on $a$ and $b$. Nonetheless, the two terms relevant for this discussion do not dependent on $a$ nor $b$, but on the sum of both. Therefore, it is irrelevant how we choose to distribute the derivatives; the result will continue to be the same, i.e. terms proportional to $a_i a^i$ will not be generated.

We can also argue how the results will change if we consider interactions with derivatives of the field in the action. For example, if we add a term of the type $N \sqrt{g} \phi^3 \Delta \phi$, the 1-loop evolution equation of $\langle \phi \rangle$ will get corrections proportional to $\langle \hat{\phi} \Delta \hat{\phi} \rangle$, $\langle \hat{\phi} \partial_i \hat{\phi} \rangle$ and $\langle \partial_i \hat{\phi} \partial^i \hat{\phi} \rangle$, besides the already known $\langle \hat{\phi}^2 \rangle$. These are calculated from the imaginary part of the Feynman propagator, taking the appropriate derivatives prior to taking the coincidence limit. This will bring down powers of $k$ in a Fourier space representation, so the second adiabatic order part of these expectation values can be expressed as
\begin{eqnarray}
\int \frac{d^4 k}{(2 \pi)^4}\, f(k^i)\, \tilde{G}_{N2}(k) ,
\end{eqnarray}
where $\tilde{G}_{N2}(k)$ is the same from Eq. \eqref{G-orders-N} and is given in the Appendix. As it happens in the calculation of Eq. \eqref{results-N2-1}, the terms proportional to $a_i a^i$ vanish before the $d^3 k$ integrals are performed. Hence, the presence of these extra  $k^i$ factors does not change that fact. Accordingly, interaction terms with derivatives of the field seem not to generate corrections with $a_i a^i$ either.

It remains to be shown whether it is true in general that Lagrangians of the form 
\begin{eqnarray}
\mathscr{L} = N\sqrt{g} \mathcal{L} ,
\label{Lagrangian-factorized} 
\end{eqnarray}
with $\mathcal{L} \neq \mathcal{L}(\partial_i N)$, do not generate quantum corrections that couple $\langle \phi \rangle$ with $a_i a^i$. On the other hand,  these corrections are generated if the classical Lagrangian contains explicit couplings of the quantum fields with  $\partial_i N$, see for instance Ref. \cite{rusos}.

\section{Conclusions}

 In this paper, we studied some aspects of Lifshitz-type field theories
 in curved spaces,  towards the main goal of assessing  the phenomenological
viability of theories with broken Lorentz invariance. We focused on two aspects: 1) the development of methods for analyzing the 
running of the constants in the Lagrangian
(and with that, the eventual recovery of the Lorentz invariance  at low energies); 2) the study of nonminimal couplings between matter fields and the spacetime metric.
 Regarding the second point, we looked at two types of terms: couplings with the spatial derivative of the lapse function (i.e., with $a_ia^i$ where $a_i=\partial_i \ln N$)
 and covariant and noncovariant couplings with the curvature scalars.

We presented a detailed study of the 1-loop renormalizability of $z=3$ self-interacting Lifshitz fields in curved spacetimes, for interactions of the form $\lambda \phi^n$.
 As the higher spatial derivatives of the propagators improve the ultraviolet behavior of the theory, the divergences appear only in the zeroth adiabatic order; that is, they
 are essentially the flat spacetime divergences. For the computation of the RG equations for the coupling constants both in the UV and in the IR, we generalized the exact RG 
methods to the Lifshitz case.  A full analysis of the running of these couplings would require us to solve a complicated system of  coupled nonlinear 
differential equations. However, a more interesting point would be to analyze the running of the constants corresponding to the kinetic terms. 
Therefore,  it
 would be worth it to extend the method of the exact RG we have implemented for the effective potential to the effective action in order to  study
the IR running of derivative couplings \cite{erge}. 
For this, one could use a generalization of 
the derivative expansion methods described in Refs. \cite{morris1,morris2}. 
This would be a very useful tool for analyzing the running of the  couplings corresponding to
 the IR-relevant terms with two (time and spatial) derivatives  and to study the eventual  emergence of a universal limiting speed.

We  also computed the second adiabatic order contributions to the mean value equation of the Lifshitz field to analyze the couplings to the curvature. We found 
that the self-interaction induces couplings to the four-dimensional  Ricci scalar, the 3-curvature and to the extrinsic curvature,   but it does not generate couplings with $a_ia^i$.
We  argued that the absence of quantum corrections involving  such derivatives of the lapse function seems to be  general, provided the Lagrangian is of  the form
given in Eq. (\ref{Lagrangian-factorized}).  

The  coupling with the four-dimensional  Ricci scalar diverges in the limit 
$\Lambda\to\infty$, as expected from the fact that it diverges in the usual theory. However, the divergence is logarithmic in $M/\Lambda$. Therefore, unless the effective 
mass $M$ is many orders of magnitude below $\Lambda$, in these theories, the coupling to the four-dimensional  Ricci scalar does not receive large quantum  corrections 
$\xi\gg 1$.  The  noncovariant couplings to the 3-curvature and the extrinsic curvature are expected since general covariance
 is broken already at the classical level. However, it is remarkable that there is a nonvanishing finite remnant even in the limit $\Lambda\to\infty$. 

 The nonminimal couplings to the curvature may have interesting consequences both in the early Universe and in astrophysics, in regions of strong
gravitational fields.  For example, there are  proposals in which the
 Higgs boson is supposed to be responsible for inflation \cite{Shaposhnikov1}. In the standard ($z=1$) scenario, the key point of this proposal is the nonminimal coupling 
of the Higgs field to gravity \cite{Shaposhnikov1}. It has been shown that $\xi\sim 10^4$ is required in order to successfully obtain inflation and a power spectrum of 
the primordial perturbations in good agreement with observational data [Higgs-inflationary models with realistic quadratic Higgs self-interaction and minimal coupling 
($\xi=0$) generates an unacceptably large power spectrum of the primordial perturbations] \cite{Shaposhnikov1}. From our results, it is clear that such
large couplings could be generated if $M\ll \Lambda$, or in the presence of a mechanism similar to that proposed in Ref. \cite{Donoghue}, in order to
enhance the logarithmic dependence.
Like the usual coupling with ${}^{(4)}R$, the noncovariant ones could also be relevant 
 in inflationary models where a Lifshitz field generates the primordial perturbations \cite{mukohyama1,mukohyama2,Qiu}.
These proposals  have been questioned
 on the ground that they might suffer from unitary problems, due to the large value of the coupling to the curvature (see, for instance,
Refs. \cite{Lerner,Burgess,Shaposhnikov2,Atkins,Gian}). Recently, in the context of Ho\v{r}ava-Lifshitz gravity, it has been shown that inflationary models with a $z=3$ ``Higgs''
field with $\xi=0$ could be constructed  \cite{Qiu}, so that one would not need to worry about preventing unitarity problems. It
 would be interesting to further analyze these kinds of models. 
 
 For a nonminimally coupled 
$z=1$ Higgs field, observational consequences in strong-gravity astrophysical environments of the curvature dependence of the mass generation 
have been analyzed in Refs. \cite{Onofrio1,Onofrio2}. As ${}^{(4)}R$ vanishes for spherically symmetric sources described by the Schwarzschild  
metric,  in order to obtain nonvanishing effects, the author assumed an unnatural coupling to the square root of the Kretschmann invariant $K_1=R_{\mu\nu\rho\sigma}R^{\mu\nu\rho\sigma}$.  For the case of $z=3$ Lifshitz fields, the interactions generate
couplings to ${}^{(3)}R$ and $K$, which do not vanish for the Schwarzschild  metric. Therefore, it is not necessary to assume a coupling
to the Kretschmann invariant. 
In Refs.  \cite{Onofrio1,Onofrio2} it was found  that,  unless one allows quite large values for the associated coupling constant to $\sqrt{K_1}$, the effects are quite far from what can be achieved with any foreseeable survey. 
Hence, we can expect a similar conclusion for the noncovariant couplings.
 
Finally, as for future work,  it would be interesting  to consider theories containing  fermions and gauge fields.
From the phenomenological point of view, in addition to  studies  of the running of the coupling constants in flat space, 
 it would  also be of  interest to  analyze the nonminimal couplings induced by quantum corrections---in particular  in the context of electrodynamics
in curved spaces---and use 
astrophysical and cosmological
 observational data to constraint the parameters of the action, generalizing the analysis of Ref. \cite{qedcs} to the case of
 noncovariant couplings to the spacetime metric.

\section*{Acknowledgements}
This work was supported by ANPCyT, CONICET and UBA. The authors thank Diego Blas, Jorge Russo and H\'ector Vucetich for interesting discussions.

\appendix

\section*{Appendix A: First and second adiabatic order expressions for the propagator}\label{apB}

In this Appendix, we provide some explicit expressions for the adiabatic expansion of the propagator. In the case of the weak gravitational field described in Sec. III, the first order in this approximation involves the following expression: 
\begin{eqnarray}
f_{k}(p) &=&\left( \frac{\tilde{h}}{2}-\delta \tilde{n} \right) (k_{0}^{2}+k_{0}p_{0}) +(p_{0}k_{i}+2k_{0}k_{i}+p_{i}k_{0})\delta \tilde{N}^{i} -\left( \frac{\tilde{h}}{2}+\delta \tilde{n} \right) \omega _{k}^{2} \notag \\
&&+\tilde{h}_{ij}k^{i}k^{j} \frac{d\omega _{k}^{2}}{dk^{2}}+\tilde{h}_{ij}\delta _{rs}k^{i}k^{j}k^{r}p^{s}\frac{d^{2}\omega _{k}^{2}}{d(k^{2})^{2}} -\left[ \left( \frac{\tilde{h}}{2}+\delta \tilde{n}%
\right) \delta _{ij}\text{ }p^{i}k^{j}-\tilde{h}^{ij}p_{i}k_{j}\right] \frac{d\omega
_{k}^{2}}{dk^{2}}  \notag \\
&&-\frac{d^{2}\omega _{k}^{2}}{d(k^{2})^{2}}\left( \frac{\delta \tilde{n}}{2}%
p^{2}k^{2}+\frac{\tilde{h}}{2}\left( \delta _{ij}k^{i}p^{j}\right) ^{2}-\frac{1}{2}%
\tilde{h}_{ij}k^{i}k^{j}p^{2}-\tilde{h}_{ij}\delta _{rs}p^{i}p^{r}k^{j}k^{s}\right)  \notag
\\
&&+\frac{d^{3}\omega _{k}^{2}}{d(k^{2})^{3}}\left( \frac{\delta \tilde{n}}{4}%
p^{2}k^{4}+\frac{2}{3}\tilde{h}_{ij}k^{i}k^{j}(\delta _{rs}k^{r}p^{s})^{2}-\frac{%
\delta \tilde{n}}{3}(\delta _{ij}k^{i}p^{j})^{2}k^{2}\right)\notag \\ 
&&-\frac{d^{3}\omega _{k}^{2}}{d(k^{2})^{3}}\left( \frac{\delta \tilde{n}}{3}%
p^{2}k^{2}\delta _{ij}k^{i}p^{j}+\frac{\tilde{h}}{3}\left( \delta
_{ij}k^{i}p^{j}\right) ^{3}-\frac{2}{3}\tilde{h}_{ij}k^{i}k^{j}p^{2}\delta
_{rs}k^{r}p^{s}-\frac{2}{3}\tilde{h}_{ij}p^{i}k^{j}(\delta
_{rs}k^{r}p^{s})^{2}\right)  \notag \\
&&-\frac{d^{3}\omega _{k}^{2}}{d(k^{2})^{3}}\left( \frac{\delta \tilde{n}}{12}%
p^{4}k^{2}+\frac{\tilde{h}}{3}p^{2}\left( \delta _{ij}k^{i}p^{j}\right) ^{2}-\frac{1%
}{6}\tilde{h}_{ij}k^{i}k^{j}p^{4}-\frac{2}{3}\tilde{h}_{ij}\delta
_{rs}p^{i}p^{r}k^{j}k^{s}p^{2}\right)  \notag \\
&&-\frac{d^{3}\omega_{k}^{2}}{d(k^{2})^{3}}\left( \frac{\tilde{h}}{12}p^{4}\delta
_{ij}k^{i}p^{j}-\frac{1}{6}\tilde{h}_{ij}k^{i}p^{j}p^{4}\right) -\frac{d^{2}\omega
_{k}^{2}}{d(k^{2})^{2}}\left( \frac{\tilde{h}}{4}p^{2}\delta _{ij}k^{i}p^{j}-\frac{1%
}{2}\tilde{h}_{ij}k^{i}p^{j}p^2\right).
\label{fkp}
\end{eqnarray}
Then, the first and second adiabatic orders for the propagator are given by
\begin{eqnarray}
[G_{\textup{F}}^{(L)}]^{AD(1)} &=& - i~\int \frac{d^3 k}{(2\pi)^3}~\Biggl\{ D^0_2(k) k_i \partial_t \delta N^i + D^0_2(k) \partial_i \left[ \left( \frac{h}{2}+\delta n \right) \delta^{ij}-h^{ij}\right] k_{j} \frac{d\omega_{k}^{2}}{dk^{2}} \notag \\
&&~~~~~ - D^0_2(k) k^{i} \partial_i h^{jk} k_{j}k_{k} \frac{d^{2}\omega _{k}^{2}}{d(k^{2})^{2}} + 2 D^2_3(k) \frac{d\omega_{k}^{2}}{dk^{2}} k^i \partial_i \left( \frac{h}{2}-\delta n \right) \notag \\
&&~~~~~ - 2 D^0_3(k) \omega_{k}^{2} \frac{d\omega_{k}^{2}}{dk^{2}} k^i \partial_i \left( \frac{h}{2}+\delta n \right) + 2 D^0_3(k) \left(\frac{d\omega_{k}^{2}}{dk^{2}} \right)^2 k^i \partial_i h^{jk} k_j k_k \notag \\
&&~~~~~ + 4 D^2_3(k) k_i \partial_t \delta N^i \Biggl\}, \label{GF-adiab-1-coinc}
\end{eqnarray}
\begin{eqnarray}
[G_{\textup{F}}^{(L)}]^{AD(2)} &=& - \int \frac{d^3 k}{(2\pi)^3} ~ \Biggl\{ \frac{1}{2} D^0_2(k) \frac{d^{2}\omega _{k}^{2}}{d(k^{2})^{2}} \Biggl( k^2 \partial^2 \delta n + k^i k^j \partial_i \partial_j h - \partial^2 h^{ij} k_i k_j - 2 k^i \partial_i \partial_j h^{jk} k_k \Biggr) \notag \\ 
&& - D^0_2(k) \frac{d^{3}\omega _{k}^{2}}{d(k^{2})^{3}} \Biggl( \frac{k^4}{4} \partial^2 \delta n + \frac{2}{3} k^i k^j \partial_i \partial_j h^{kl} k_k k_l - \frac{k^2}{3} k^i k^j \partial_i \partial_j \delta n \Biggr) \notag \\
&& + D^0_3(k) \frac{d\omega _{k}^{2}}{dk^{2}} \Biggl[ -2k^i \partial_i \partial_t \delta N^j k_j - 2\frac{d\omega _{k}^{2}}{dk^{2}} k^i k^j \partial_i \partial_j \left( \frac{h}{2}+\delta n \right) + 2\frac{d\omega _{k}^{2}}{dk^{2}} k^i k_j \partial_i \partial_k h^{jk} \notag \\
&& ~~~~~ + 2\frac{d^{2}\omega _{k}^{2}}{d(k^{2})^{2}} k^i k^j \partial_i \partial_j h^{kl} k_k k_l - \omega_k^2 \partial^2 \left( \frac{h}{2}+\delta n \right) + \frac{d\omega _{k}^{2}}{dk^{2}} \partial^2 h^{ij} k_i k_j \Biggr] \notag \\
&& + D^0_3(k) \Biggl[ -2 \omega _{k}^{2} \frac{d^{2}\omega _{k}^{2}}{d(k^{2})^{2}} k^i k^j \partial_i \partial_j \left( \frac{h}{2}+\delta n \right) + 2 \frac{d\omega _{k}^{2}}{dk^{2}} \frac{d^{2}\omega _{k}^{2}}{d(k^{2})^{2}} k^i k^j \partial_i \partial_j h^{kl} k_k k_l \notag \\
&&~~~~~ + \omega_k^2 \partial^2_t \left( \frac{h}{2}+\delta n \right) - \frac{d\omega _{k}^{2}}{dk^{2}} \partial^2_t h^{ij} k_i k_j \Biggr] + D^2_3(k) \Biggl[ -3 \partial^2_t \left( \frac{h}{2}-\delta n \right) \notag \\
&& + 2\partial_t \partial_i \delta N^i + 2 \frac{d^{2}\omega _{k}^{2}}{d(k^{2})^{2}} k^i k^j \partial_i \partial_j \left( \frac{h}{2}-\delta n \right) + \frac{d\omega _{k}^{2}}{dk^{2}} \partial^2 \left( \frac{h}{2}-\delta n \right) \Biggr] \notag \\
&&+ 4 D^0_4(k) \left( \frac{d\omega _{k}^{2}}{dk^{2}} \right)^2 k^i k^j \partial_i \partial_j \Biggl[ \left( \frac{h}{2}+\delta n \right) \omega_k^2 - h^{kl} k_k k_l \frac{d\omega _{k}^{2}}{dk^{2}} \Biggr] \notag \\
&&-4 D^2_4(k) \Biggl[ \left( \frac{d\omega _{k}^{2}}{dk^{2}} \right)^2 k^i k^j \partial_i \partial_j \left( \frac{h}{2}-\delta n \right) + 4 \frac{d\omega _{k}^{2}}{dk^{2}} k^i \partial_i \partial_t \delta N^j k_j - \omega_k^2 \partial^2_t \left( \frac{h}{2}+\delta n \right) \notag\\
&&~~~~~ + \frac{d\omega _{k}^{2}}{dk^{2}} \partial^2_t h^{ij} k_i k_j \Biggr] -4 D^4_4(k) \partial^2_t \left( \frac{h}{2}-\delta n \right) \Biggr\} . \label{GF-adiab-2-coinc}
\end{eqnarray}

On the other hand, for the metric considered in Sec. VI A, the expansion of the propagator in derivatives of the lapse function $N$ is

\begin{eqnarray}
\tilde{G}_{N0}(k) &=& -\frac{1}{N (\omega_k^2 - \tilde{k}_0^2)} ,\\
\tilde{G}_{N1}(k) &=& - i \frac{(\omega_k^2 + 3\tilde{k}_0^2)}{N (\omega_k^2 - \tilde{k}_0^2)^3} \left( \frac{d \omega_k^2}{d k^2} \right) k^i a_i ,\\
\tilde{G}_{N2}(k) &=& - \frac{1}{N} \Biggl\{ \frac{(3\tilde{k}_0^4 + 4\tilde{k}_0^2 \omega_k^2 + \omega_k^4)}{(\omega_k^2 - \tilde{k}_0^2)^4} \left( 2 \frac{d^2 \omega_k^2}{d (k^2)^2} k^i k^j + \frac{d \omega_k^2}{d k^2} \delta^{ij} \right) +\frac{(\omega_k^2 + 3\tilde{k}_0^2)}{(\omega_k^2 - \tilde{k}_0^2)^4} \left( \frac{d \omega_k^2}{d k^2} \right)^2 k^i k^j \notag \\ 
&&- \frac{(20\tilde{k}_0^4 + 24\tilde{k}_0^2 \omega_k^2 + 4\omega_k^4)}{(\omega_k^2 - \tilde{k}_0^2)^5} \left( \frac{d \omega_k^2}{d k^2} \right)^2 k^i k^j + \frac{2\tilde{k}_0^2}{(\omega_k^2 - \tilde{k}_0^2)^3} \frac{d \omega_k^2}{d k^2} \delta^{ij} \notag \\
&&- \frac{8\tilde{k}_0^2}{(\omega_k^2 - \tilde{k}_0^2)^4} \left( \frac{d \omega_k^2}{d k^2} \right)^2 k^i k^j + \frac{4\tilde{k}_0^2}{(\omega_k^2 - \tilde{k}_0^2)^3} \frac{d^2 \omega_k^2}{d (k^2)^2} k^i k^j \Biggr\} a_i a_j \notag \\
&&- \frac{1}{N} \Biggl\{ - \frac{(\omega_k^2 + \tilde{k}_0^2)}{(\omega_k^2 - \tilde{k}_0^2)^3} \left[ \left( \frac{d \omega_k^2}{d k^2} \right) \delta^{ij} + 2 \left( \frac{d^2 \omega_k^2}{d (k^2)^2} \right) k^i k^j \right] + \frac{4(\omega_k^2 + \tilde{k}_0^2)}{(\omega_k^2 - \tilde{k}_0^2)^4} \left( \frac{d \omega_k^2}{d k^2} \right)^2 k^i k^j \notag \\
&&- \frac{2}{(\omega_k^2 - \tilde{k}_0^2)^3} \left( \frac{d \omega_k^2}{d k^2} \right)^2 k^i k^j + \frac{k^2}{2} \frac{1}{(\omega_k^2 - \tilde{k}_0^2)^2} \left( \frac{d^2 \omega_k^2}{d (k^2)^2} - \frac{k^2}{2} \frac{d^3 \omega_k^2}{d (k^2)^3} \right) \delta^{ij} \notag \\
&&+ \frac{k^2}{3} \frac{1}{(\omega_k^2 - \tilde{k}_0^2)^2} \frac{d^3 \omega_k^2}{d (k^2)^3} k^i k^j \Biggr\} b_{ij} ,\label{GN2-Fourier}
\end{eqnarray}
where $\tilde{k}_0 \equiv k_0/N$.

\end{document}